%% file: main.tex
\documentclass[conference]{IEEEtran}
\IEEEoverridecommandlockouts
\usepackage{cite}
\usepackage{amsmath,amssymb,amsfonts}
\usepackage{graphicx}
\usepackage{subcaption}
\usepackage{textcomp}
\usepackage{color}
\usepackage{booktabs}
\usepackage[normalem]{ulem}
\usepackage{enumitem}
\usepackage{multirow}
\usepackage{url}
\usepackage{hyperref}
\usepackage{algorithm}
\usepackage{algorithmicx}
\usepackage{algpseudocode}
\usepackage[table,xcdraw]{xcolor} 

\definecolor{brightturquoise}{rgb}{0.85, 1, 1}

\def\BibTeX{{\rm B\kern-.05em{\sc i\kern-.025em b}\kern-.08em
    T\kern-.1667em\lower.7ex\hbox{E}\kern-.125emX}}
\begin{document}

\title{LCS-CTC: Leveraging Soft Alignments to Enhance Phonetic Transcription Robustness}

% \author{Anonymous ASRU 2025 submission}
\author{%
  \IEEEauthorblockN{%
    \parbox{\linewidth}{\centering
      Zongli Ye\IEEEauthorrefmark{1}, %
      Jiachen Lian\IEEEauthorrefmark{2}\thanks{Project Lead / Corresponding author: \texttt{jiachenlian@berkeley.edu}}, %
    Akshaj Gupta\IEEEauthorrefmark{2}, Xuanru Zhou\IEEEauthorrefmark{1}, Haodong Li\IEEEauthorrefmark{3}, Krish Patel\IEEEauthorrefmark{2},\\
      Hwi Joo Park\IEEEauthorrefmark{2}, Dingkun Zhou\IEEEauthorrefmark{4}, Chenxu Guo\IEEEauthorrefmark{1}, Shuhe Li\IEEEauthorrefmark{1}, Sam Wang\IEEEauthorrefmark{2}, Iris Zhou\IEEEauthorrefmark{2}, Cheol Jun Cho\IEEEauthorrefmark{2}, \\Zoe Ezzes\IEEEauthorrefmark{5},
      Jet M.J. Vonk\IEEEauthorrefmark{5}, Brittany T. Morin\IEEEauthorrefmark{5}, Rian Bogley\IEEEauthorrefmark{5}, Lisa Wauters\IEEEauthorrefmark{5}, Zachary A. Miller\IEEEauthorrefmark{5},\\
      Maria Luisa Gorno-Tempini\IEEEauthorrefmark{5}, Gopala Anumanchipalli\IEEEauthorrefmark{2}%
    }
  }\\[0.1em]   
  \IEEEauthorblockA{\normalsize  
    \IEEEauthorrefmark{1}Zhejiang University \quad
    \IEEEauthorrefmark{2}UC Berkeley \quad
    \IEEEauthorrefmark{3}SUSTech \quad
    \IEEEauthorrefmark{4}SCUT \quad
    \IEEEauthorrefmark{5}UCSF
  }
}

\maketitle

\input{abstract}
\input{introduction}

\input{methods}

\input{experiments}
\input{conclusion}
% \input{extra}

% \input{References}
% \clearpage
\bibliographystyle{IEEEtran}
\bibliography{refs}{}
% \bibliography{IEEEabrv, refs}
\end{document}

%% file: abstract.tex
\begin{abstract}
Phonetic speech transcription is crucial for fine-grained linguistic analysis and downstream speech applications. While Connectionist Temporal Classification (CTC) is a widely used approach for such tasks due to its efficiency, it often falls short in recognition performance, especially under unclear and nonfluent speech. In this work, we propose LCS-CTC, a two-stage framework for phoneme-level speech recognition that combines a similarity-aware local alignment algorithm with a constrained CTC training objective. By predicting fine-grained frame-phoneme cost matrices and applying a modified Longest Common Subsequence (LCS) algorithm, our method identifies high-confidence alignment zones which are used to constrain the CTC decoding path space, thereby reducing overfitting and improving generalization ability, which enables both robust recognition and text-free forced alignment. Experiments on both LibriSpeech and PPA demonstrate that LCS-CTC consistently outperforms vanilla CTC baselines, suggesting its potential to unify phoneme modeling across fluent and non-fluent speech. 
\end{abstract}


%% file: introduction.tex
\section{Introduction}

Within-word variation in human speech poses challenges for ASR systems~\cite{radford2022whisper, zhang2023google-usm, pratap2024scaling-mms, puvvada2024less-canary-nvidia, nvidia2025parakeet} that focus solely on word-level transcription. Yet, modeling subword structures like syllables or phonemes is essential for applications in language learning and clinical analysis~\cite{kheir-etal-2023-automatic-assessment-review,truong04_icall-acoustuc-phonetic,lee2016language-capt,cordella2024connected-fluency1,fontan2023automatically-fluency2,gordon1998fluency-fluency3,gordon2022clinicians-fluency4,metu2023evaluating-fluency5,ssdm, lian2024ssdm2.0, zhou2024yolostutterendtoendregionwisespeech, zhou2024stutter, zhou2024timetokensbenchmarkingendtoend}. This work focuses on phoneme transcription—a particularly challenging yet critical task.

Several factors contribute to the difficulty of phoneme-level transcription. First, unlike word-level targets, phoneme annotations are inherently non-deterministic due to accent, allophony and other context-dependent realizations~\cite{choi2025leveragingallophonyselfsupervisedspeech}. Second, existing training objectives, such as CTC~\cite{graves2006connectionist-ctc} and attention-based models~\cite{chorowski2015attention-asr}, often introduce alignment noise. For example, CTC tends to produce overly peaked label distributions~\cite{zeyer2021does-ctc-peaky}, which result in inaccurate timing and error-prone alignments. Attention-based models, on the other hand, may hallucinate transcriptions~\cite{koenecke2024careless}—an issue that is particularly problematic in clinical applications where precision is critical. Third, most existing remedies are designed for fluent speech and demonstrate limited effectiveness on non-fluent speech. For example, several approaches~\cite{tian2022bayes-ctc1, huang2024less-ctc2, yao2024cr-ctc3} enhance CTC-based alignment by introducing heuristic constraints on the prior. However, these heuristic alignments have been shown to be incompatible with non-fluent speech~\cite{lian2023unconstrained-udm, lian-anumanchipalli-2024-towards-hudm,ssdm, lian2024ssdm2.0, ye2025seamlessalignment-neurallcs}. 

Given the aforementioned challenges, a key question arises: \textit{Can we design a phoneme transcription objective that performs robustly simultaneously on both fluent and non-fluent speech, while also producing reliable and accurately aligned outputs suitable for clinical applications?} 

A key insight arises from treating \textit{phoneme transcription as a speech production process}, particularly when speakers are prompted to read reference text. Fluent speech results from accurate production, while deviations introduce non-fluent speech. For example, when given the phoneme sequence “IH N S ER T,” a speaker might produce “IH [S] N S ER [AH] T,” introducing insertions on “IH” and “ER.” This yields an alignment such as “IH–(IH,S), N–(N), S–(S), ER–(ER,AH), T–(T)” (label–spoken). The core idea is that the \textit{loss function should account only for effective speech-text alignments}. In the case of “IH–(IH,S),” frames aligned to the extraneous “S” should be excluded when computing the loss for “IH.” Without such errors, the method reduces to the standard ASR objective. This alignment strategy follows the Longest Common Subsequence (LCS) principle~\cite{hirschberg1977algorithms-lcs1}, which was recently shown effective for non-fluent speech~\cite{ssdm}.
Building on this intuition, we apply the LCS algorithm online to identify effective alignments, which are then used to regularize the vanilla CTC~\cite{graves2006connectionist-ctc} paths, as shown in Fig.~\ref{fig:pipeline}. We refer to this method as \textbf{LCS-CTC}. Unlike previous methods that introduce heuristics to regularize CTC alignments~\cite{tian2022bayes-ctc1, huang2024less-ctc2, yao2024cr-ctc3}, \textit{LCS-CTC is grounded in human speech production, interpretable, and clinically meaningful}.

Our LCS-CTC framework consists of two main components: a phoneme-aware alignment module based on a modified Longest Common Subsequence (LCS) algorithm, and a constrained CTC training objective that incorporates the resulting alignment masks. The alignment module first predicts a frame-phoneme cost matrix using a lightweight neural network trained with weak supervision. This matrix encodes pairwise acoustic-phonetic similarity and is used to derive a partial alignment path via a similarity-aware LCS dynamic programming procedure. The resulting alignment mask identifies high-confidence frame-phoneme pairs and is used in two complementary ways during training: (1) it anchors specific frames to phoneme labels with cross-entropy supervision, and (2) it restricts the CTC path space via masked emission probabilities. Together, these two components enable robust phoneme recognition by combining interpretable alignment guidance with flexible sequence modeling.

We evaluate our method on both fluent (LibriSpeech~\cite{7178964}) and non-fluent corpora, including PPA speech~\cite{gorno2011classification-ppa} and the largest existing simulated non-fluent corpus, \textit{LLM\_dys}~\cite{zhang2025analysisevaluationsyntheticdata}. Results show that LCS-CTC consistently outperforms vanilla CTC across all metrics, including (weighted) phoneme error rate and duration-aware alignment accuracy. Notably, while LCS-CTC is inspired by non-fluent speech, it also provides stronger regularization for fluent speech. Although we have not yet explored its scalability across data types, linguistic diversity, or domains, the method shows strong potential and may serve as a foundation for universal phoneme recognition. We have open-sourced our model and checkpoints at \href{https://github.com/Auroraaa86/LCS-CTC}{\texttt{https://github.com/Auroraaa86/LCS-CTC}}

%% file: methods.tex
\section{Methods}

\begin{figure}[t]
    \centering
    \begin{subfigure}{\columnwidth}
        \centering
        \includegraphics[width=0.9\textwidth]{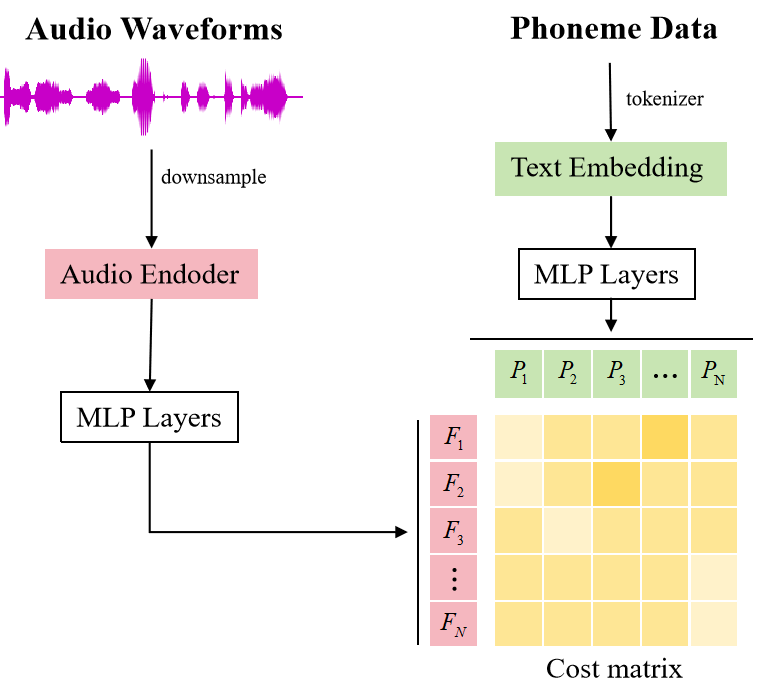}
        \caption{}
        \label{fig:sub1}
    \end{subfigure}
    
    \vspace{1em}
    \begin{subfigure}{\columnwidth}
        \centering
        \includegraphics[width=1.0\textwidth]{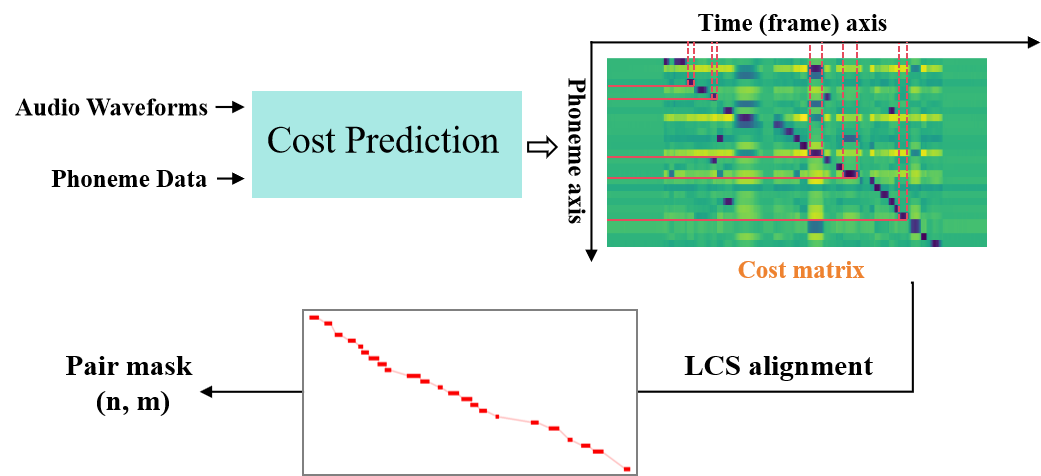}
        \caption{}
        \label{fig:sub2}
    \end{subfigure}
    
    \caption{(a) shows the structure of cost matrix learning model. (b) shows the process from predicted cost matrix to alignment mask: we use cost matrix to construct the dynamic programming matrix and implement modified LCS algorithm to realize a local wise alignment, the dark red dots indicate credible local alignment regions.}
    \label{fig:combined}
    \vspace{-1.8em}
\end{figure}

\begin{figure*}[t]
  \centering
  \includegraphics[width=0.8\textwidth]{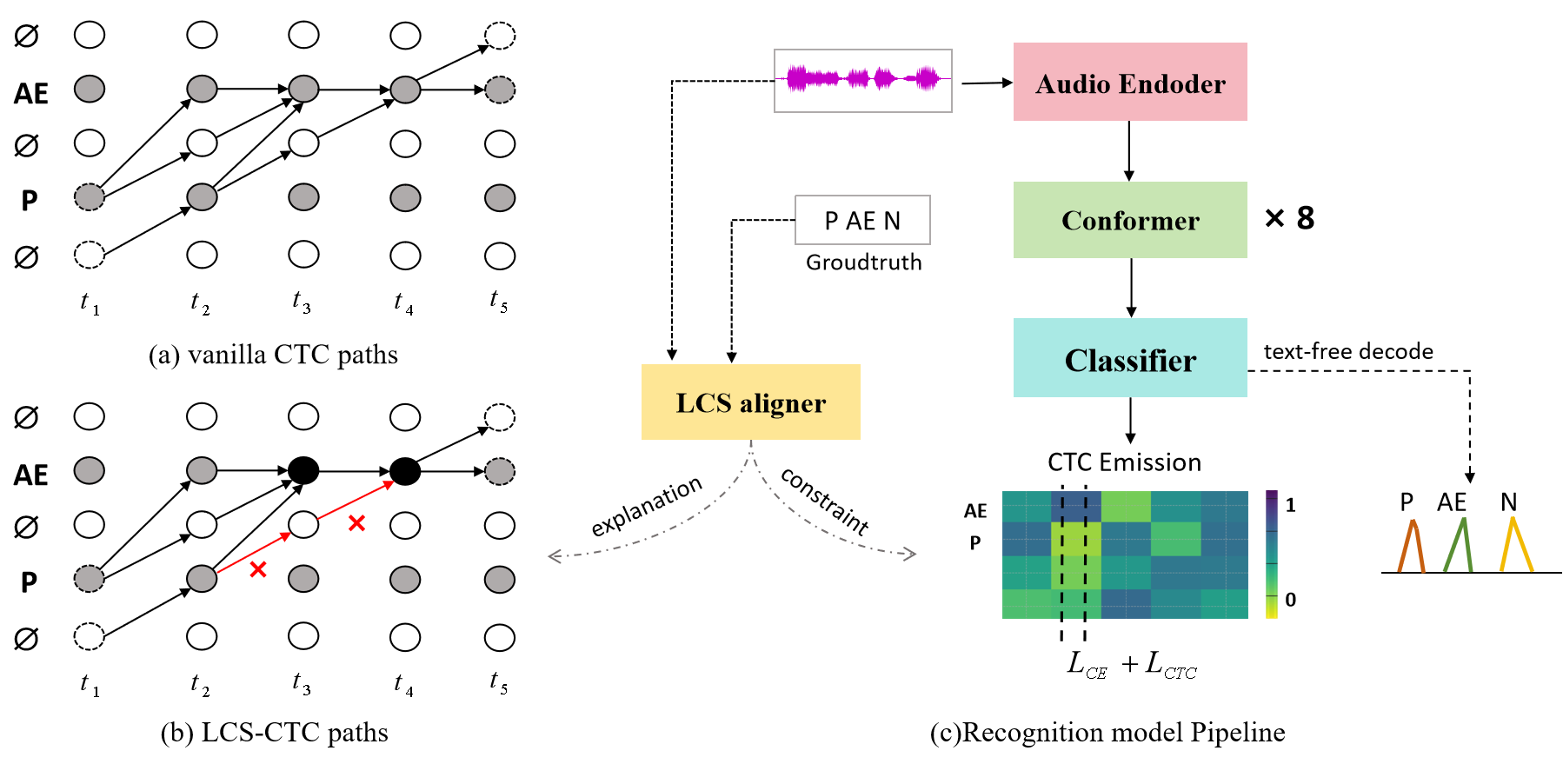}
  \caption{\textit{Overview of LCS-CTC framework}: (a) Vanilla CTC allows all valid alignment paths that collapse to the target sequence, which can result in peaky distributions.  
(b) Our LCS-CTC introduces alignment constraints from the LCS-based aligner (black nodes), which eliminate implausible paths (red crosses) by fixing "high-confidence" frame-phoneme matches.  
(c) The overall training pipeline of LCS-CTC. The LCS aligner uses ground-truth phonemes and predicted emissions to generate alignment masks, which are then used both as supervision (via $\mathcal{L}_{\text{CE}}$) and constraint (via masked $\mathcal{L}_{\text{CTC}}$) during recognition model training.}
  \label{fig:pipeline}
      \vspace{-11pt}
\end{figure*}

\begin{algorithm}[h]
\caption{LCS-based Sequence Alignment}\label{alg1}
\footnotesize
\begin{algorithmic}[1]
\Statex \textbf{Input:} 
    $Label$ (phoneme labels), 
    $C$ (cost matrix), 
    $tol$ (tolerance),
    $phn\_sim$ (similarity dictionary)
\Statex \textbf{Output:} Alignment mask

\Procedure{Align}{$Label$, $C$, $tol$, $phn\_sim$}
    \State $valid\_res \gets \emptyset$
    \For{$(i,j) \in [0,n) \times [0,m)$}
        \State $phn1 \gets Label[i]$, $phn2 \gets Label[k]$ \Comment{$k=\arg\min C[:,j]$}
        \State $sim \gets phn\_sim[(phn1,phn2)]$ \Comment{Lookup similarity}
        \If{$C[i][j] \leq (1-sim) \times tol$} \Comment{Similarity-adjusted threshold}
            \State $valid\_res \gets valid\_res \cup \{(i,j), (j,i)\}$
        \EndIf
    \EndFor
    
    \State $dp \gets \text{zeros}(n+1,m+1)$
    \For{$i \gets 1$ to $n$}
        \For{$j \gets 1$ to $m$}
            \If{$(i-1,j-1) \in valid\_res$}
                \State $dp[i][j] \gets \max(dp[i-1][j-1]+1, dp[i][j-1])$
                \For{$k \gets j+1$ to $m$} \Comment{Propagate valid matches}
                    \State $dp[i][k] \gets dp[i][k-1] + \mathbb{I}_{(i-1,k-1) \in valid\_res}$
                \EndFor
\Statex \quad The indicator function $\mathbb{I}_{(i{-}1,k{-}1) \in \textit{valid\_res}}$ ensures that alignment credit is propagated horizontally across frames that are all considered valid matches to phoneme $i{-}1$.
            \Else
                \State $dp[i][j] \gets \max(dp[i-1][j], dp[i][j-1])$
            \EndIf
        \EndFor
    \EndFor

    \State $path \gets$ \Call{Traceback}{$dp$} \Comment{Recover alignment path from DP}
    \State \Return $\text{MaskFromPath}(path)$
\EndProcedure
\end{algorithmic}
\end{algorithm}

The LCS-CTC framework comprises two stages: local alignment extraction via similarity-aware LCS, and constrained CTC training using the resulting alignment masks. This design aims to improve ASR robustness by enforcing phonemically plausible supervision while preserving temporal flexibility.

\subsection{Similarity-Guided LCS Alignment}
\label{sec:lcs-align}

In this section, we propose a novel method for frame-wise speech-text alignment inspired by the Longest Common Subsequence (LCS) algorithm~\cite{hirschberg1977algorithms-lcs1}. Unlike traditional forced alignment approaches that enforce strict one-to-one alignment between each speech frame and a corresponding phoneme label, our method relaxes this assumption by focusing only on “highly confident" frame-label pairs. This partial alignment strategy better reflects the acoustic variability in natural speech, such as ambiguous onsets and offsets, where the phonetic boundaries are inherently fuzzy.

\paragraph{Overview} Our method consists of two key stages: (1) predicting a fine-grained cost matrix between speech frames and phoneme labels using a neural model, and (2) applying a similarity-aware dynamic programming alignment inspired by LCS. The goal is to identify optimal frame-label matches that reflect both acoustic evidence and phonetic similarity. The process of our method is shown in Fig.~\ref{fig:combined}

\paragraph{Cost Matrix Learning} Given a speech utterance segmented into $m$ frames and its corresponding phoneme sequence of length $n$, we train a neural model to predict a cost matrix $C \in \mathbb{R}^{n \times m}$ where $C[i,j]$ estimates the alignment cost between the $i$-th phoneme and the $j$-th frame. The model is trained using a subset of VCTK~\cite{yamagishi2019cstr-vctk} data, for which accurate frame-level alignments are obtained via Montreal Forced Aligner (MFA)~\cite{mcauliffe17_interspeech} and manually corrected annotations.

To construct target labels for supervision, we incorporate a pre-defined phoneme similarity function $s: \mathcal{P_\text{cmu}} \times \mathcal{P_\text{cmu}} \rightarrow [0,1]$, where $\mathcal{P_\text{cmu}}$ denotes the set of CMU phonemes. This similarity is computed based on a set of eight articulatory features, including phoneme type (vowel/consonant), vowel length, height, frontness, lip rounding, consonant manner, place of articulation, and voicing. Let the ground-truth aligned phoneme for frame $t_j$ be $p^*_j$; then the target cost for matching phoneme $p_i$ to frame $t_j$ is defined as:
\[
\tilde{C}[i,j] =
\begin{cases}
0, & \text{if } p_i = p^*_j, \\
1 - s(p_i, p^*_j), & \text{otherwise}.
\end{cases}
\]

To account for temporal uncertainty at phoneme boundaries (e.g., transitional blur at start or end of a phoneme), we further apply a Gaussian edge attenuation to the ground-truth aligned region.Finally, the cost matrix labels $\tilde{C}$ are normalized across the time axis via softmax. For convenience, we denote $s(p_i,p_j)$ as the similarity values after softmax.

To learn this cost matrix label, we design a model that jointly encodes both the audio signal and the target phoneme sequence. The audio input is first passed through a pretrained audio encoder (e.g., Wav2Vec2~\cite{baevski2020wav2vec20frameworkselfsupervised}), producing a sequence of frame-level acoustic embeddings. Simultaneously, the phoneme sequence is mapped into dense embeddings. Each of the two modalities is then processed by separate MLPs to project them into a shared hidden space. The projected frame and phoneme features are combined via negative dot product, followed by a softmax over the time axis. The resulting matrix is then trained to match the cost matrix labels using a Kullback-Leibler (KL) divergence loss. The structure of the model is illustrated in Fig.~\ref{fig:sub1}.

\paragraph{Similarity-Aware LCS Alignment} Once the model yields a predicted cost matrix $C$, we proceed to perform a similarity-informed dynamic alignment. The alignment algorithm identifies all phoneme-frame pairs $(i,j)$ which the predicted cost $C[i,j]$ falls below a similarity-adjusted threshold $\tau_{i,j} = (1 - s(p_i, p_j)) \cdot \text{tol}$, where $\text{tol}$ is a global tolerance hyperparameter, and we set tol = 1 in our work, the selection process is shown in Section~\ref{abl}. This yields a set of valid match candidates:
\[
\begin{aligned}
valid\_res &= \{(i,j) \mid C[i,j] \leq (1 - s(p_i, p_k)) \cdot \text{tol}\}, \\
&\quad \text{with } k = \arg\min_{i} C[i,j].
\end{aligned}
\]

The implementation of the algorithm is shown in Algorithm~\ref{alg1}. It outputs binary alignment mask $M^{'} \in \{0,1\}^{n \times m}$ derived from the alignment path where $M^{'}[i,j] = 1$ if phoneme $p_i$ aligns with frame $t_j$. This design allows integration into traditional ASR pipelines which we will talk below.

By integrating phonetic similarity into the cost prediction and alignment stages, our method enables more robust alignment in the presence of acoustic ambiguity or phoneme confusability. The process from cost matrix to alignment mask can be seen in Fig.~\ref{fig:sub2}.

\subsection{Constrained LCS-CTC for ASR Training}

\subsubsection{Review on Connectionist Temporal Classification (CTC)}

The CTC loss has been widely adopted in ASR systems for its ability to model frame-to-label alignments with length mismatch. Given an input sequence of acoustic features $X = \{x_1, x_2, \dots, x_T\}$ and a target label sequence $Y = \{y_1, y_2, \dots, y_L\}$, CTC computes the loss by marginalizing over all possible frame-to-label alignments $\pi \in \mathcal{B}^{-1}(Y)$ where $\mathcal{B}$ is the many-to-one mapping function that removes blank tokens and repeated labels. Vanilla CTC loss is defined as:

\begin{equation*}
\mathcal{L}_{\text{CTC}} = -\log \sum_{\pi \in \mathcal{B}^{-1}(Y)} P(\pi \mid X)
\end{equation*}

 While flexible, vanilla CTC suffers from the \emph{peaky distribution} problem, where probability mass tends to concentrate on a few frames, leading to unreliable and non-robust alignments. Moreover, it treats all frame-label alignments as equally likely during training, regardless of their phonetic plausibility, which can harm generalization in downstream recognition tasks.

\subsubsection{LCS-CTC: Constrained CTC Training with Alignment Masks}

To enhance the robustness and generalization ability of the recognition model trained by CTC, we propose a novel CTC variant named \textbf{LCS-CTC}, which integrates frame-phoneme alignment masks derived from similarity-aware LCS algorithm (as introduced in Section~\ref{sec:lcs-align}). The central idea is to apply the LCS-derived alignment mask as a constraint over the CTC emission probabilities, thereby pruning unlikely frame-to-phoneme associations during training. In our recognition model, the input audio signal is first encoded by a pretrained speech encoder, followed by a stack of 8 Conformer layers~\cite{gulati2020conformer} and an MLP-based classifier head that produces the final emission probabilities. Since the emission matrix $P \in \mathbb{R}^{L \times T}$ represents the frame-wise posterior over the entire phoneme vocabulary $\mathcal{P_\text{cmu}}$ of size $L$, we transform the original alignment mask $M^{'}$(of shape $n \times T$ for a target sequence of length $n$) into a binary matrix $M \in \{0,1\}^{L \times T}$, where $M[i,j] = 1$ indicates that frame $t_j$ is constrained to emit phoneme $p_i$. This transformation ensures compatibility between the alignment constraint and the CTC output space.

The masked emission $\tilde{P}$ is computed by applying the alignment mask $M$ to the original CTC emission $P$:

% \begin{equation*}
% \tilde{P}[i,j] = \frac{P[i,j] \cdot M[i,j]  + \epsilon}{\sum_{k} P[i,k] \cdot M[i,k]}
% \end{equation*}

\begin{equation*}
\tilde{P}[i,j] =
\begin{cases}
\displaystyle\frac{P[i,j] \cdot M[i,j] + \epsilon}{\sum_{l} P[l,j] \cdot M[l,j] + \epsilon}, & \text{if } \sum_{l} M[l,j] > 0 \\
P[i,j], & \text{otherwise}
\end{cases}
\end{equation*}

where $\epsilon$ is a small constant to ensure numerical stability. This operation selectively normalizes emission probabilities at frames that are constrained by the LCS-derived alignment mask, while leaving unconstrained frames unchanged. 

We then define a hybrid training objective that supervises the model in two complementary ways. First, for the high-confidence alignment regions specified by the binary mask $M$, we apply cross-entropy loss directly on the original emission $P$ to provide strong local supervision. Second, we apply vanilla CTC loss over the full sequence using the masked emission $\tilde{P}$, which softly restricts the CTC decoding space during sequence-level training. Crucially, by anchoring certain frames to fixed phoneme labels through $M$, our method effectively eliminates implausible alignment paths and narrows the search space, thereby improving CTC training stability and generalization. The details are illustrated in Fig.~\ref{fig:pipeline}.

The final loss is computed as:
\begin{equation*}
\mathcal{L}_{\text{LCS-CTC}} = \lambda \cdot \mathcal{L}_{\text{CE}}(P \odot M, Y) + (1 - \lambda) \cdot \mathcal{L}_{\text{CTC}}(\tilde{P}, Y)
\end{equation*}

where $\odot$ denotes element-wise masking. Here, $\mathcal{L}_{\text{CE}}$ denotes standard cross-entropy loss evaluated only on masked positions, while $\lambda \in [0,1]$ is a weighting factor balancing between strict alignment enforcement and flexible CTC decoding. We set $\lambda = 0.5$ in our work. This hybrid objective ensures that the model is supervised on phonemically plausible alignments while retaining the benefits of sequence-level flexibility.

%% file: experiments.tex
\section{Experiments}

\begin{table*}[ht]
\centering
\begin{tabular}{l|l|cccc|ccc|c}
\toprule
\multirow{2}{*}{\textbf{Model}} & \multirow{2}{*}{\textbf{Method}} & \multicolumn{4}{c|}{\textbf{LibriSpeech}} & \multicolumn{3}{c|}{\textbf{PPA}} &\multirow{2}{*}{\textbf{LLM\_dys}}\\
& & \textit{test-clean} & \textit{test-other} & \textit{dev-clean} & \textit{dev-other} & \textit{nfvppa} & \textit{lvppa} & \textit{svppa} & \\
\midrule
\multirow{3}{*}{\textbf{Wav2Vec2.0-L}} 
& CTC      & 16.16 & 21.04 & 15.41 & 20.50 & 41.04 & 36.63 & 15.78 & 14.52 \\
& CE-CTC   & 16.84 & 21.16 & 15.73 & 20.68 & 38.89 & 38.75 & 17.18 & 14.75 \\
& LCS-CTC\textit{(ours)}  & \textbf{16.09} & \textbf{20.94} & \textbf{15.10} & \textbf{19.70} & \textbf{38.14} & \textbf{32.04} & \textbf{15.01} & \textbf{13.89} \\
\midrule
\multirow{3}{*}{\textbf{HuBERT-L }} 
& CTC      & 15.00 & 18.16 & 14.12 & 17.18 & 32.70 & 18.34 & 13.32 & 12.13 \\
& CE-CTC   & 14.83 & 17.75 & 14.02 & 17.07 & 34.45 & 19.06 & 15.24 & 12.94 \\
& LCS-CTC\textit{(ours)}  & \textbf{14.75} & \textbf{17.35} & \textbf{11.78} & \textbf{16.72} & \textbf{32.02} & \textbf{17.71} & \textbf{13.30} & \textbf{11.85} \\
\midrule
\multirow{3}{*}{\textbf{WavLM-L}} 
& CTC      & 12.62 & 15.01 & 13.92 & 14.99 & 29.88 & 20.33 & 14.14 & 12.13 \\
& CE-CTC   & 12.31 & 14.73 & 12.83 & 14.90 & 30.34 & 22.15 & 15.05 & 11.30 \\
& LCS-CTC\textit{(ours)}  & \textbf{12.09} & \textbf{14.45} & \textbf{12.41} & \textbf{14.73} & \textbf{29.51} & \textbf{18.72} & \textbf{13.83} & \textbf{11.22} \\
\bottomrule
\end{tabular}
\caption{PER(\%) performance of three methods on LibriSpeech, PPA, and LLM\_dys dataset.}
\label{tab:per-results}
 \vspace{-5pt}
\end{table*}
\begin{table*}[ht]
\centering
\begin{tabular}{l|l|cccc|ccc|c}
\toprule
\multirow{2}{*}{\textbf{Model}} & \multirow{2}{*}{\textbf{Method}} & \multicolumn{4}{c|}{\textbf{Librispeech}} & \multicolumn{3}{c|}{\textbf{PPA}} & \multirow{2}{*}{\textbf{LLM\_dys}} \\
 & & \textit{test-clean} & \textit{test-other} & \textit{dev-clean} & \textit{dev-other} & \textit{nfvppa} & \textit{lvppa} & \textit{svppa} & \\
\midrule
\multirow{3}{*}{\textbf{WavLM-L}} 
 & CTC & 10.75 & 13.41 & 11.96 & 12.07 & 23.14 & 11.46 & 11.70 & 6.26 \\
 & CE-CTC & 10.73 & 13.41 & 11.70 & 12.14 & 23.55 & 13.05 & 12.94 & 5.71 \\
 & LCS-CTC\textit{(ours)} & \textbf{10.49} & \textbf{12.85} & \textbf{11.56} & \textbf{11.97} & \textbf{22.75} & \textbf{10.92} & \textbf{11.00} & \textbf{5.64} \\
\bottomrule
\end{tabular}
\caption{WPER(\%) performance of three methods on LibriSpeech, PPA, and LLM\_dys dataset.}
\label{tab:wper-results}
 \vspace{-5pt}
\end{table*}

\begin{table*}[ht]
\centering
\begin{tabular}{c|c|cccc|c}
\toprule
\multirow{2}{*}{\textbf{Model}} & \multirow{2}{*}{\textbf{Method}} 
& \multicolumn{4}{c|}{\textbf{Librispeech}} & \multirow{2}{*}{\textbf{LLM\_dys}} \\
& & \textit{test-clean} & \textit{test-other} & \textit{dev-clean} & \textit{dev-other} & \\
\midrule
\multirow{2}{*}{\textbf{WavLM-L}} 
& CTC      & 146.6 & 113.5 & 115.8 & 131.3 & 95.1 \\
& LCS-CTC\textit{(ours)}  & \textbf{115.8} & \textbf{94.5} & \textbf{94.9} & \textbf{102.9} & \textbf{76.7} \\
\bottomrule
\end{tabular}
\caption{BL(ms) performance of different methods on LibriSpeech and LLM\_dys dataset.}
\label{tab:boundary-loss}
 \vspace{-5pt}
\end{table*}

\begin{table*}[ht]
\centering
\begin{tabular}{c|c|cccc|c}
\toprule
\multirow{2}{*}{\textbf{Model}} & \multirow{2}{*}{\textbf{Method}} 
& \multicolumn{4}{c|}{\textbf{Librispeech}} & \multirow{2}{*}{\textbf{LLM\_dys}} \\
& & \textit{test-clean} & \textit{test-other} & \textit{dev-clean} & \textit{dev-other} & \\
\midrule
\multirow{2}{*}{\textbf{WavLM-L}} 
& CTC      & 22.4 & 14.8& 19.7 & 28.0 & 17.9 \\
& LCS-CTC\textit{(ours)}  &\textbf{16.0}  & \textbf{10.8} & \textbf{13.1} & \textbf{21.9} & \textbf{12.7} \\
\bottomrule
\end{tabular}
\caption{ARL performance of different methods on LibriSpeech and LLM\_dys dataset.}
\label{tab:arl-loss}
    \vspace{-10pt}
\end{table*}

\subsection{Datasets}
We use \textbf{VCTK}~\cite{yamagishi2019cstr-vctk} dataset which includes 109 native English speakers with accented speech to train our cost matrix learning model and phoneme recognition model, and we conduct experiments on three ASR datasets to evaluate our proposed LCS-CTC's performance: (1) \textbf{Librispeech}~\cite{7178964}, which contains 1000 hours of English speech, we use its \textit{dev} and \textit{test} subsets for evaluation. (2) \textbf{PPA
Speech}~\cite{gorno2011classification-ppa}, it is collected in collaboration with clinical experts and includes recordings from 38 participants diagnosed with
Primary Progressive Aphasia (PPA). Participants were asked to
read the ”grandfather passage,” resulting in approximately one
hour of speech in total. (3) \textbf{LLM\_dys}, as mentioned in~\cite{zhang2025analysisevaluationsyntheticdata}, We used a large language model (claude-3-5-sonnet-20241022~\cite{anthropic2024claude}) to create a synthetic speech dataset. Given clean text and its CMU/IPA sequences, the model was prompted to insert dysfluencies naturally. This produced dysfluent IPA sequences for VITS-based speech synthesis~\cite{kim2021conditionalvariationalautoencoderadversarial}, as well as CMU sequences labeled with dysfluencies for evaluation. 
\subsection{Implementation details}
In our main experiments, we split the VCTK dataset into four partitions with a ratio of 3:1:5:1. Specifically, 30\% of the data is used for training the cost matrix learning model. These samples are annotated with MFA and manual refinement, mentioned in Section~\ref{sec:lcs-align}. An additional 10\% is held out as the validation set for this stage.

For training the phoneme recognition model based on our proposed LCS-CTC, we use 50\% of the data for training and 10\% for validation. All experiments are conducted on an NVIDIA A6000 GPU. We use a learning rate of $1\text{e}{-5}$ and a batch size of 1 for all training stages. No weight decay is applied during optimization.

\subsection{Evaluation Metrics}
\subsubsection{Phoneme Error Rate (PER)}
Phoneme Error Rate is a common metric for evaluating phoneme-level recognition performance. It is defined as the Levenshtein distance between the predicted and reference phoneme sequences, normalized by the length of the reference. The metric accounts for substitution, insertion, and deletion errors, and reflects the overall transcription accuracy at the phonetic level.
\subsubsection{Weighted Phonetic Error Rate (WPER)}Unlike PER which assumes all phonemes are equally distant, WPER takes phonetic variation into account by incorporating phoneme similarity. This addresses the shortcomings of traditional error metrics and offers a more accurate evaluation of recognition performance. Refer to~\cite{guo2025dysfluentwfstframeworkzeroshot}, we define WPER as
\[
\text{WPER} = \frac{\sum_{(p_t,p_s)} (1 - s(p_t,p_s)) + D + I}{N}
\]

where \( p_t, p_s \) denote target and substitute phonemes, \( s \) is their similarity score defined in \ref{sec:lcs-align}, 
and \( N \) is the reference phoneme sequence length. \( D \) and \( I \) represent deletion 
and insertion errors respectively. For substitutions, phoneme similarity is evaluated first,
followed by error penalties.
\subsubsection{Boundary Loss (BL)}
Boundary Loss (BL) quantifies the alignment quality between predicted and ground-truth phoneme boundaries. Specifically, it is computed as the average absolute deviation between the predicted and reference onset/offset times of each phoneme. A lower BL reflects more accurate temporal alignment.
\subsubsection{Articulatory Reconstruction Loss (ARL)}
To assess phoneme-level articulatory consistency, we introduce ARL. It measures the L2 distance between articulatory embeddings reconstructed from predicted phonemes and those extracted from the original audio using an acoustic-to-articulatory inversion (AAI) model~\cite{Cho_2024}. Lower ARL indicates better articulatory plausibility of the recognition output.

\subsection{Results and Discussion}

\begin{figure*}[t]
    \centering
    \includegraphics[width=15.0cm]{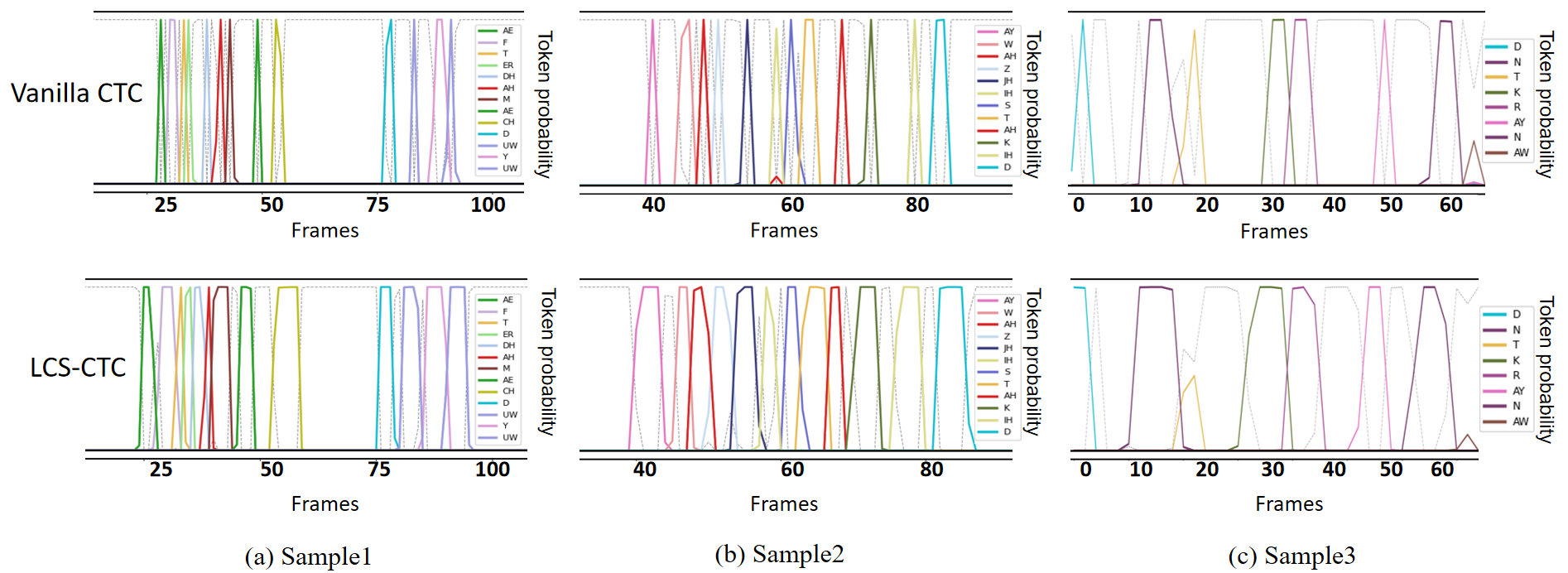}
    \caption{Visualization of token emission probabilities for vanilla CTC and our proposed LCS-CTC on three randomly selected samples from the LibriSpeech and PPA. The gray dashed lines indicate the blank token. Compared to vanilla CTC, LCS-CTC produces smoother and more distributed token activations, characterized by more consistent repetitions of non-blank tokens.}
    \label{fig:emission}
     \vspace{-10pt}
\end{figure*}

Table~\ref{tab:per-results} presents the PER of different training methods, including vanilla CTC and our proposed LCS-CTC—applied to three common speech encoders: Wav2Vec2.0~\cite{baevski2020wav2vec20frameworkselfsupervised}, HuBERT~\cite{hsu2021hubertselfsupervisedspeechrepresentation}, and WavLM~\cite{Chen_2022}. Across all models and datasets, LCS-CTC outperforms vanilla CTC on both clean (e.g., test-clean, dev-clean) and challenging conditions (e.g., test-other, lvppa, LLM\_dys), demonstrating strong generalization.

We also compare with CE-CTC, a variant that combines CTC with frame-level cross-entropy loss derived from ground-truth alignments. While CE-CTC introduces additional frame-level supervision, it does not consistently outperform vanilla CTC, especially on more challenging datasets such as PPA and LLM\_dys. This suggests that relying on full-frame labels does not always lead to better generalization, and may even introduce noise under dysfluent or noisy speech conditions. In contrast, LCS-CTC achieves competitive or superior performance without relying on frame-level labels at training time, making it more practical and scalable.

Furthermore, we report WPER in Table~\ref{tab:wper-results} using the best-performing encoder WavLM. LCS-CTC again shows the best performance on all subsets. These results confirm that our method leads to more accurate and robust phoneme recognition across various speech conditions.

We use Boundary Loss (BL) to measure alignment quality. As shown in Table~\ref{tab:boundary-loss}, LCS-CTC significantly reduces BL compared to vanilla CTC, indicating more accurate and temporally consistent alignments. This is further supported by the emission visualizations in Fig.~\ref{fig:emission}, where LCS-CTC produces smoother and less peaky token distributions. Such smoother emission behavior enhances generalization and likely contributes to the improved recognition performance observed earlier. Moreover, our model can also serve as a text-free phoneme-level aligner. The results in Table~\ref{tab:arl-loss} demonstrate that LCS-CTC yields consistently lower articulatory reconstruction loss compared to vanilla CTC. This suggests that the phoneme sequences predicted by LCS-CTC better preserve articulatory smoothness and continuity, leading to embeddings that more closely match the natural trajectories derived from the speech signal.

\subsection{Ablation experiments}\label{abl}
\vspace{-5pt}
\begin{figure}[ht]
  \centering
  \includegraphics[width=0.9\columnwidth]{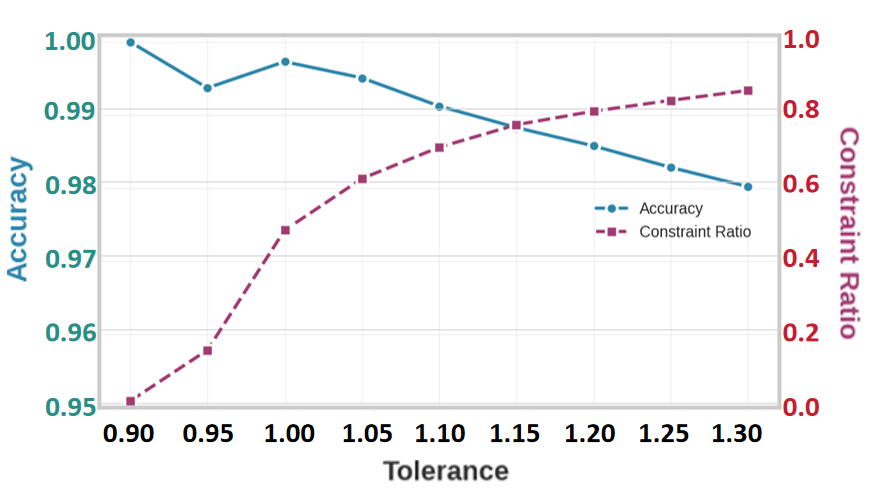}
  \caption{Effect of \textit{tol} on alignment accuracy and constrained region ratio.}
  \label{fig:tol}
  \vspace{-5pt}
\end{figure}
We perform an ablation study on the tolerance parameter $\mathit{tol}$ in the LCS alignment algorithm, which controls the threshold for accepting frame-phoneme matches based on predicted cost and phoneme similarity. Smaller values of $\mathit{tol}$ impose stricter alignment conditions, leading to higher alignment precision but fewer constrained regions. This may reduce the effectiveness of the frame-level supervision in our LCS-CTC. In contrast, larger $\mathit{tol}$ values include more frames but risk introducing noisy or imprecise alignments, which can weaken training. We evaluate values in the range $\{0.9-1.3\}$ and observe that $\mathit{tol} = 1.0$ provides the best balance between alignment accuracy and coverage, and is thus adopted in all main experiments.

%% file: conclusion.tex
\vspace{-0.3em}
\section{Conclusions and Limitations}

In this work, we propose LCS-CTC which demonstrates strong performance across all evaluation metrics for both fluent and non-fluent speech, and, critically, yields outputs that are more clinically interpretable. Nonetheless, several limitations remain. The current framework is trained on clean speech from the VCTK corpus, which is insufficient to fully assess its generalizability to more diverse, noisy, or pathological speech. Future work will focus on expanding both the training and evaluation datasets to encompass a broader range of conditions. LCS-CTC operates explicitly as a forced aligner, maintaining transparency comparable to traditional HMM-based systems such as MFA~\cite{mcauliffe17_interspeech}. A promising direction for future research is the development of a neural variant~\cite{pratap2024scaling-mms, zhu2022phone-w2v2-alignment, li2022neufa, seamless2025joint} that retains interpretability while achieving performance on par with established HMM-based methods. Another challenge involves the inherent ambiguity of phonetic labels. To address this, we plan to incorporate phonetic similarity modeling~\cite{zhou2025phonetic-error-detection} and WFST-based decoding~\cite{guo2025dysfluentwfstframeworkzeroshot, li2025k} strategies. Moreover, given that phonemes are intrinsically articulatory in nature, we aim to integrate articulatory priors~\cite{cho2024jstsp-sparc, ssdm, lian22bcsnmf, lian2023factor} into the label space to mitigate alignment instability—particularly in disordered or atypical speech.

\vspace{-0.3em}
\section{Acknowledgements}
Thanks for support from UC Noyce Initiative, Society of Hellman Fellows, NIH/NIDCD, and the Schwab Innovation fund. Many thanks to Alexei Baevski for the early-stage discussions and insightful idea brainstorming.